\documentclass[prd,onecolumn,nopacs,nofootinbib]{revtex4}

\usepackage{amsmath}
\usepackage{graphicx}
\usepackage{amsfonts}
\usepackage{latexsym}
\usepackage{bbold}
\usepackage{wasysym}
\usepackage{calligra}
\usepackage{float}
\usepackage{ulem}
\usepackage{inputenc}
\usepackage{xspace}
\usepackage{url}
\usepackage{epstopdf}
\usepackage{tikz}
\usepackage{amsthm}

\usepackage[framemethod=TikZ]{mdframed}
\usepackage{lipsum}

\newcommand{\be}{\begin{equation}}
\newcommand{\ee}{\end{equation}}
\newcommand{\beq} {\begin{equation}}
\newcommand{\eeq} {\end{equation}}
\newcommand{\ba}{\begin{eqnarray}}
\newcommand{\ea}{\end{eqnarray}}

\newtheorem{mydef}{Definition}
\newtheorem{corollary}{Corollary}

\newtheorem{theo}{Theorem}
\newenvironment{ftheo}
{\begin{mdframed}\begin{theo}}
		{\end{theo}\end{mdframed}}

\begin{document}

	\title{Riemann Tensor and Gauss-Bonnet density in Metric-Affine Cosmology}
	
	\author{Damianos Iosifidis}
	\affiliation{Institute of Theoretical Physics, Department of Physics
		Aristotle University of Thessaloniki, 54124 Thessaloniki, Greece}
	\email{diosifid@auth.gr}
	
	\date{\today}
	\begin{abstract}
		
We analytically derive the covariant form of the Riemann (curvature) tensor for homogeneous Metric-Affine Cosmologies. That is, we present, in a Cosmological setting, the most general covariant form of the full Riemann tensor including also its non-Riemannian pieces which are associated to spacetime torsion and non-metricity. Having done so we also compute a list of the curvature tensor by-products such as Ricci tensor, homothetic curvature, Ricci scalar, Einstein tensor etc. Finally we derive the generalized Metric-Affine version of the usual Gauss-Bonnet density in this background and demonstrate how under certain circumstances the latter represents a total derivative term.
		
	\end{abstract}
	
	\maketitle
	
	\allowdisplaybreaks
	
	
	\tableofcontents
	
	\section{Introduction}
	\label{intro}
	
	Non-Riemannian effects (i.e. torsion and non-metricity) are believed to play a crucial role in the early time Cosmological evolution \cite{puetzfeld2005prospects}. These non-Riemannian contributions which are associated with the microstructure of matter \cite{hehl1976hypermomentum,hehl1997ahadronic,obukhov1993hyperfluid,Iosifidis:2020gth} appear naturally in the so-called Metric-Affine Gravity (MAG) formulation \cite{hehl1995metric,iosifidis2019metric} where the metric and the affine connection are regarded as independent variables with the latter possessing both torsion and non-metricity. There are many characteristics of MAG that make it an attractive and promising Theory of Gravity. For instance, when written in the language of exterior differential forms, it can be seen to be a gauge Theory of Gravity \cite{hehl1995metric}. In addition, it takes into account the microscopic characteristics of matter such as spin, dilation and shear and also offers appealing features of quantization \cite{percacci2020towards}. Possible experimental effects of these non-Riemannian effects were discussed in \cite{kostelecky2008constraints,foster2017constraints}. 
	
	This Metric-Affine framework has received a renewed attention over the last years \cite{iosifidis2018raychaudhuri,iosifidis2019exactly,iosifidis2019scale,vitagliano2011dynamics,sotiriou2007metric,capozziello2010metric,percacci2020new,Jimenez:2020dpn,BeltranJimenez:2019acz,aoki2019scalar,Cabral:2020fax,Ariwahjoedi:2020wmo,Yang:2021fjy,helpin2020metric,bahamonde2020new} especially in regards to its cosmological applications \cite{iosifidis2020cosmic,Iosifidis:2020gth,Iosifidis:2021iuw,jimenez2016spacetimes,beltran2017modified,kranas2019friedmann,barragan2009bouncing,shimada2019metric,kubota2021cosmological,Mikura:2020qhc,mikura2021minimal}. It therefore becomes apparent that it is of great importance to have a complete setup for both the geometrical and the matter structures on this non-Riemannian Cosmology. Regarding the latter point, we had earlier developed in \cite{Iosifidis:2020gth} the generalized fluid notion that sources the non-Riemannian degrees of freedom in a Cosmological setup. We named this perfect fluid generalization (where the microscopic characteristics of matter are taken into account) the Perfect Cosmological Hyperfluid \cite{Iosifidis:2020gth}\footnote{Quite recently we have generalized this hyperfluid notion even further, even outside of the realm of Cosmology \cite{iosifidis2021perfect}.}. There it was shown that this kind of fluid is the appropriate one for the description of the matter sources in Metric-Affine Cosmology, since it sources all the non-Riemannian degrees of freedom that are allowed by the Cosmological Principle.  Having the complete set for the sources one also needs the associated compatible set of geometrical objects in this Metric-Affine Cosmological setting. These are torsion, non-metricity and curvature, the three fundamental geometrical objects of the non-Riemannian geometry \cite{eisenhart2012non}. The covariant expressions for  both torsion and non-metricity, compatible with the Cosmological Principle, have also been presented in \cite{Iosifidis:2020gth}. Therefore, the only remaining missing link is the covariant expression for the Riemann (curvature) tensor that respects both isotropy and homogeneity. It is this gap that a part of this work here intends to fill. It goes without saying that having the exact covariant form of the Riemann tensor serves as a very helpful tool for studying the Cosmological aspects of Metric-Affine Gravity and especially Theories that are constructed by the latter, such as $f(R)$ \cite{sotiriou2007metric,olmo2011palatini} or certain generalizations of it \cite{delhom2019ricci}.
	 Of course, having the covariant form of the Riemann tensor one could then go on and compute several important by-products of the latter such as Ricci tensor, Ricci scalar, Riemann dual etc. We list several of these quantities after the derivation of the Riemann tensor. In addition, one highly interesting quantity one can construct, when having this covariant form of the Riemann tensor, is the well-known Gauss-Bonnet density. Using a certain generalization of the aforementioned density for non-Riemannian spaces, we derive its form for a homogeneous torsionful and non-metric Cosmology. As expected we find that when the full non-metricity is present the Gauss-Bonnet density fails to be an exact differential \cite{babourova1997pontryagin,janssen2019topological}. It is also known that for the restricted class of Cartan-Weyl spaces the generalized Gauss-Bonnet density is indeed topological. In our Cosmological setting we extend this result here for a slightly more general class of non-metricity, that is for geometries that generalize the Cartan-Weyl structure. 
	 
	 The paper is organized as follows. Firstly, we fix notation and definitions and go through the basic geometric and Cosmological setup. We then derive the covariant form of the Riemann tensor for homogeneous and isotropic Cosmologies with the inclusion of both torsion and non-metricity. Continuing, we present a list of the forms of some important quantities that appear in actual applications (Ricci tensor and scalar, Einstein tensor, totally antisymmetric Riemann etc). Finally with the use of the previously derived Riemann tensor we obtain the form of the generalized Gauss-Bonnet density for non-Riemannian Cosmologies. We confirm that the latter is not a total divergence in general  and then  prove that for a certain geometric subsector of the full non-metricity tensor the latter is a total derivative.

		\section{The setup}
	Let us consider a generalized $n-dim$ non-Riemannian Geometry, where the space apart from curvature possesses also torsion and non-metricity. Following Schouten \cite{schouten1954ricci} we shall refer to this  general $n-dim$ manifold endowed with a metric and an independent affine connection as an \underline{$L_{n}$ space}. We will use the definitions and notations of \cite{Iosifidis:2020gth} so we will go through the basic setup rather briefly here and refer the reader to \cite{Iosifidis:2020gth} for more details. We consider a non-Riemannian manifold endowed with a metric $g_{\mu\nu}$ and an independent affine connection $\Gamma^{\lambda}_{\;\;\;\mu\nu}$. We define the curvature, torsion and non-metricity tensors according to
	\beq
	R^{\mu}_{\;\;\;\nu\alpha\beta}:= 2\partial_{[\alpha}\Gamma^{\mu}_{\;\;\;|\nu|\beta]}+2\Gamma^{\mu}_{\;\;\;\rho[\alpha}\Gamma^{\rho}_{\;\;\;|\nu|\beta]} \label{R}
	\eeq
	\beq
	S_{\mu\nu}^{\;\;\;\lambda}:=\Gamma^{\lambda}_{\;\;\;[\mu\nu]}
	\eeq
	\beq
	Q_{\alpha\mu\nu}:=- \nabla_{\alpha}g_{\mu\nu}
	\eeq
	and the deviation of the affine connection $\Gamma^{\lambda}_{\;\;\;\mu\nu}$ from the Levi-Civita one defines the distortion tensor \cite{schouten1954ricci}
	\begin{gather}
	N^{\lambda}_{\;\;\;\;\mu\nu}:=\Gamma^{\lambda}_{\;\;\;\mu\nu}-\widetilde{\Gamma}^{\lambda}_{\;\;\;\mu\nu}=
	\frac{1}{2}g^{\alpha\lambda}(Q_{\mu\nu\alpha}+Q_{\nu\alpha\mu}-Q_{\alpha\mu\nu}) -g^{\alpha\lambda}(S_{\alpha\mu\nu}+S_{\alpha\nu\mu}-S_{\mu\nu\alpha}) \label{N}
	\end{gather}
	where $\widetilde{\Gamma}^{\lambda}_{\;\;\;\mu\nu}$ is the usual Levi-Civita connection calculated only by the metric and its first derivatives. Once the distortion is given, torsion and non-metricity can be quite easily computed through (see for instance \cite{iosifidis2019metric})
	\beq
	S_{\mu\nu\alpha}=N_{\alpha[\mu\nu]}\;\;,\;\;\; Q_{\nu\alpha\mu}=2 N_{(\alpha\mu)\nu} \label{N1}
	\eeq
	Out of torsion we can construct a vector as well as a pseudo-vector. Our definitions for the torsion vector and pseudo-vector are
	\beq
	S_{\mu}:=S_{\mu\lambda}^{\;\;\;\;\lambda} \;\;, \;\;\;
	\tilde{S}_{\mu}:=\epsilon_{\mu\alpha\beta\gamma}S^{\alpha\beta\gamma} 
	\eeq
	respectively. Note that the former is defined for any dimension and the latter only for $n=4$. Continuing with non-metricity, we define the Weyl  and the second non-metricity vector according to
	\beq
	Q_{\alpha}:=Q_{\alpha\mu\nu}g^{\mu\nu}\;,\;\; \tilde{Q}_{\nu}=Q_{\alpha\mu\nu}g^{\alpha\mu}
	\eeq
	Finally,  from the Riemann tensor we can construct the three contractions
	\beq
	R_{\nu\beta}:=R^{\mu}_{\;\;\nu\mu\beta}	
	\eeq
\beq
\hat{R}_{\alpha\beta}:=R^{\mu}_{\;\;\mu\alpha\beta}	
\eeq
	\beq
	\breve{R}^{\mu}_{\;\;\beta}:=R^{\mu}_{\;\;\nu\alpha\beta}	g^{\nu\alpha}
	\eeq
	
	The first one is the usual Ricci tensor (which is not symmetric now), the second one is the homothetic curvature and the last one is oftentimes referred to as the co-Ricci tensor. Note that the first two aforementioned tensors can be formed without the use of any metric while for the latter a metric is needed.
Finally, let us mention that by virtue of ($\ref{N}$) each quantity can be split into its Riemannian part (i.e. computed with respect to the Levi-Civita connection) plus non-Riemannian contributions. For instance inserting the connection decomposition ($\ref{N}$) into the definition ($\ref{R}$) we obtain for the Riemann tensor \footnote{Quantities with\;  $\widetilde{}$\;  will always denote Riemannian parts  unless otherwise stated.} 
\beq
{R^\mu}_{\nu \alpha \beta} = \widetilde{R}^\mu_{\phantom{\mu} \nu \alpha \beta} + 2 \widetilde{\nabla}_{[\alpha} {N^\mu}_{|\nu|\beta]} + 2 {N^\mu}_{\lambda|\alpha} {N^\lambda}_{|\nu|\beta]} \,, \label{decomp}
\eeq
The last decomposition will be essential for the proof of the first Theorem that we give in the sequel.
	\section{Cosmology with Torsion/Non-metricity}
	Let us consider an FLRW Cosmology, with the usual Robertson-Walker line element
	\beq
	ds^{2}=-dt^{2}+a^{2}(t)\delta_{ij}dx^{i}dx^{j} \label{metric}
	\eeq
	where 	$i,j=1,2,...,n-1$ and $a(t)$ is as usual the scale factor of the Universe. As usual we will denote the Hubble parameter as $H:=\dot{a}/a$. In addition let $u^{\mu}$ represent the normalized $n$-velocity field of a given fluid which in co-moving coordinates is expressed as $u^{\mu}=\delta^{\mu}_{0}=(1,0,0,...,0)$. Accordingly we define the projection tensor in the usual manner
	\beq
	h_{\mu\nu}:=g_{\mu\nu}+u_{\mu}u_{\nu}
	\eeq
	projecting objects on the space orthogonal to $u^{\mu}$. In such a highly symmetric spacetime, torsion has only $2$ (or $1$ for $n\neq 4$)  degrees of freedom \cite{tsamparlis1979cosmological} and non-metricity contributes $3$ \cite{minkevich1998isotropic}. In \cite{Iosifidis:2020gth} we showed that the most general covariant forms of the torsion and non-metricity tensors, in such a spacetime, are given by
	\beq
	S_{\mu\nu\alpha}^{(n)}=2u_{[\mu}h_{\nu]\alpha}\Phi(t)+\epsilon_{\mu\nu\alpha\rho}u^{\rho}P(t)\delta_{n,4} \label{Scosm}
	\eeq
	\beq
	Q_{\alpha\mu\nu}=A(t)u_{\alpha}h_{\mu\nu}+B(t) h_{\alpha(\mu}u_{\nu)}+C(t)u_{\alpha}u_{\mu}u_{\nu}  \label{Qcosm}
	\eeq
	with the five functions $\Phi,P,A,B,C$ representing  the non-Riemannian Cosmological effects. Let us note that in the above expression for non-metricity the Weyl non-metricity\footnote{Recall that the Weyl non-metricity is given by $Q_{\alpha\mu\nu}=A_{\alpha}g_{\mu\nu}$.} corresponds to the special choice $B=0$ and $A=-C$. As a result the above form of non-metricity is more general than Weyl.
	
	Going back to the geometrical notions, in this case the distortion tensor is given by \cite{Iosifidis:2020gth}
	\begin{gather}
	N_{\alpha\mu\nu}^{(n)}=X(t)u_{\alpha}h_{\mu\nu}+Y(t)u_{\mu}h_{\alpha\nu}+Z(t)u_{\nu}h_{\alpha\mu}
	+V(t)u_{\alpha}u_{\mu}u_{\nu} +\epsilon_{\alpha\mu\nu\lambda}u^{\lambda}W(t)\delta_{n,4}
	\end{gather}	
	where the functions $X(t),Y(t),Z(t),V(t),W(t)$ are linearly related to $\Phi(t),P(t),A(t),B(t),C(t)$ according to
		\beq
	2(X+Y)=B \;, \;\; 2Z=A\;, \;\; 2V=C \;, \;\; 2\Phi =Y-Z\;, \;\; P = W	
	\eeq
	or inverting them
	\beq
	W=P \;, \;\; V=C/2 \;, \;\; Z=A/2	
	\eeq
	\beq
	Y=2\Phi +\frac{A}{2}	\;\;, \;\;\;
	X=\frac{B}{2}- 2 \Phi -\frac{A}{2} \label{XY}
	\eeq
	 as can be trivially checked by employing 	($\ref{N1}$). As a result, the affine connection of the non-Riemannian FLRW spacetime reads
	\beq
\Gamma^{\lambda}_{\;\;\;\mu\nu}=\widetilde{\Gamma}^{\lambda}_{\;\;\;\mu\nu}+	X(t)u^{\lambda}h_{\mu\nu}+Y(t)u_{\mu}h^{\lambda}_{\;\;\nu}+Z(t)u_{\nu}h^{\lambda}_{\;\;\mu}
	+V(t)u^{\lambda}u_{\mu}u_{\nu} +\epsilon^{\lambda}_{\;\;\mu\nu\rho}u^{\rho}W(t)\delta_{n,4} \label{connect}
	\eeq
	and recall that the non-vanishing components of the Levi-Civita connection are in this case
		\beq
	\widetilde{\Gamma}^{0}_{\;\; ij}=\widetilde{\Gamma}^{0}_{\;\; ji}=\dot{a}a \delta_{ij}=H g_{ij}
\;, \;\;
	\widetilde{\Gamma}^{i}_{\;\; j0}=\widetilde{\Gamma}^{i}_{\;\; 0j}=\frac{\dot{a}}{a}\delta^{i}_{\;\;j}=H\delta^{i}_{\;j}
	\eeq
	Now, as it is clear from the above discussion the pseudo-scalar mode $W$ (or P equivalently) exists only in $4$ dimensions and this is why the Kronecker delta appears. However,  in order to save some space and time  we will omit the inclusion of $\delta_{n,4}$, keeping in mind that $W$ vanishes for $n\neq 4$.

	Finally, let us mention that in $4$ dimensions one usually defines the three dimensional Levi-Civita form (see for instance \cite{ellis2012relativistic,tsagas2008relativistic})
	\beq
	\varepsilon_{\mu\nu\lambda}:=\varepsilon_{\mu\nu\lambda\rho}u^{\rho}
	\eeq
	living on the orthogonal tangent planes. Note that by construction $	\varepsilon_{\mu\nu\lambda}u^{\mu}=0$ confirming the fact that the latter is purely spatial. The above expression will be useful in our subsequent discussion.
	Let us close this section with one definition on which the rest of our analysis will lie upon.
	\begin{mydef}
		The metric ($\ref{metric}$) along with the connection ($\ref{connect}$) will constitute a \underline{Cosmological $L_{n}$ space}.
	\end{mydef}
	
\section{Riemann Tensor for Metric-Affine Cosmologies}
In this section we analytically  derive  the covariant expression of the Riemann tensor for a Cosmological $L_{n}$ and subsequently apply this result and compute the form of various objects constructed out of the Riemann tensor such as Ricci tensor, homothetic curvature, Ricci scalar etc.  

Let us start with the the covariant expression of the Riemann tensor which we state and prove below.

\begin{ftheo}
	Consider a Cosmological $L_{n}$ as given by the metric ($\ref{metric}$) and the connection (\ref{connect}). Then, for this geometry the \emph{Riemann (curvature) tensor} takes the covariant form
	\begin{gather}
	R_{\mu\nu\alpha\beta}	=\Big[ \dot{X}+H(X+Z+V)+X(Z+V)+(\dot{H}+H^{2})\Big] 2 h_{\nu[\alpha}u_{\beta]}u_{\mu}\nonumber \\
	+\Big[ \dot{Y}+H(Y+Z+V)-Y(Z+V)-(\dot{H}+H^{2})\Big]2 h_{\mu[\alpha}u_{\beta]}u_{\nu}\nonumber \\
	+\Big[ H(X-Y)-X Y-W^{2}+H^{2}\Big] 2 h_{\mu[\alpha}h_{\beta]\nu}\nonumber\\
	+\Big[ \dot{W}-W(H+2 X)\Big]\varepsilon_{\alpha\beta\nu}u_{\mu}-\Big[ \dot{W}-W(H-2 Y)\Big]\varepsilon_{\alpha\beta\mu}u_{\nu}-(\dot{W}+H W)\varepsilon_{\mu\nu\alpha\beta} \label{Riemann}
	\end{gather}
	This is the most general form of the Riemann tensor that respects spatial isotropy as well as homogeneity.
\end{ftheo}
\begin{proof}
	We start by the Riemann tensor decomposition $(\ref{decomp})$ and compute its parts piece by piece. For the metric $(\ref{metric})$ the Riemannian part is trivially calculated to be
		\beq
	\widetilde{R}_{\mu\nu\alpha\beta}=2 H^{2}h_{\mu[\alpha}h_{\beta]\nu}+4 (\dot{H}+H^{2})u_{[\mu}h_{\nu][\alpha}u_{\beta]}
	\eeq
	Next, using also the connection form ($\ref{connect}$), we compute
	
	\begin{gather}
	\widetilde{\nabla}_{\alpha}N^{\mu}_{\;\;\;\nu\beta}=h_{\nu\beta}\Big( -\dot{X}u^{\mu}u_{\alpha}+H X h^{\mu}_{\;\;\alpha}\Big)+H X u^{\mu}\Big( h_{\alpha\nu}u_{\beta}+h_{\alpha\beta}u_{\nu} \Big) \nonumber \\
	+h^{\mu}_{\;\;\beta}\Big( -\dot{Y}u_{\alpha} u_{\nu}+H Y h_{\alpha\nu}\Big)+ H Y u_{\nu} \Big( h_{\alpha}^{\;\;\mu}u_{\beta}+h_{\alpha\beta}u^{\mu} \Big) \nonumber \\
	+h^{\mu}_{\;\;\nu}\Big( -\dot{Z}u_{\alpha}u_{\beta}+H Z h_{\alpha\beta}\Big) + H Z u_{\beta}\Big( h_{\alpha}^{\;\;\mu}u_{\nu}+h_{\alpha\nu}u^{\mu}\Big) \nonumber \\
	+u^{\mu}u_{\nu}\Big( - \dot{V}u_{\alpha}u_{\beta}+H V h_{\alpha\beta}\Big) +H V u_{\beta}\Big( h_{\alpha}^{\;\;\mu}u_{\nu}+h_{\alpha\nu}u^{\mu}\Big) \nonumber \\
	+\varepsilon^{\mu}_{\;\;\nu\rho\beta}\Big( \dot{W}u^{\rho}u_{\alpha}- H W h_{\alpha}^{\;\;\rho} \Big)
	\end{gather}
	and antisymmetrizing in $\alpha,\beta$ it follows that
	\begin{gather}
	\widetilde{\nabla}_{[\alpha}N^{\mu}_{\;\;\;|\nu|\beta]}=\Big( \dot{X}+H(X+Z+V)\Big) h_{\nu[\alpha}u_{\beta]}u^{\mu}+H(Y-X)h_{\nu[\alpha}h_{\beta]}^{\;\;\mu}+\Big( \dot{Y}+H(Y+Z+V)\Big) h^{\mu}_{\;\;[\alpha}u_{\beta]}u_{\nu}\nonumber \\
	+\varepsilon^{\mu}_{\;\;\nu\rho[\beta}(\dot{W}u_{\alpha]}u^{\rho}-H W h_{\alpha]}^{\;\;\rho})
	\end{gather}
	Continuing, we compute
		\begin{gather}
	N^{\mu}_{\;\;\rho[\alpha}N^{\rho}_{\;\;\;|\nu|\beta]}= X (Z+V) u^{\mu}h_{\nu[\alpha}u_{\beta]}+X W u^{\mu}h^{\rho}_{\;\;[\alpha}\varepsilon_{\beta]\nu\rho\lambda}u^{\lambda}-X Y h^{\mu}_{\;\;[\alpha}h_{\beta]\nu}-Y(Z+ V) h^{\mu}_{\;\;[\alpha}u_{\beta]}u_{\nu} \nonumber \\
	+Z W h^{\mu\rho}u_{[\alpha}\varepsilon_{\beta]\rho\nu\lambda}u^{\lambda} +Y W h^{\rho}_{\;\;[\beta}\varepsilon_{\alpha]\rho}^{\;\;\;\;\;\mu\kappa}u_{\kappa}u_{\nu}-W Z \varepsilon^{\mu}_{\;\;\rho\kappa[\alpha}u_{\beta]}u^{\kappa}h^{\rho}_{\;\;\nu}+W^{2}\varepsilon^{\mu}_{\;\;\rho\kappa[\alpha}\varepsilon^{\rho}_{\;\;\beta]\nu\lambda}u^{\kappa}u^{\lambda}
	\end{gather}
	Substituting all the above into ($\ref{decomp}$) after some algebra we find 
		\begin{gather}
	R^{\mu}_{\;\;\nu\alpha\beta}	=\Big[ \dot{X}+H(X+Z+V)+X(Z+V)+(\dot{H}+H^{2})\Big] 2 h_{\nu[\alpha}u_{\beta]}u^{\mu}\nonumber \\
	+\Big[ \dot{Y}+H(Y+Z+V)-Y(Z+V)-(\dot{H}+H^{2})\Big]2 h^{\mu}_{\;\;[\alpha}u_{\beta]}u_{\nu}\nonumber \\
	+\Big[ H(X-Y)-X Y-W^{2}+H^{2}\Big] 2 h^{\mu}_{\;\;[\alpha}h_{\beta]\nu}-2XW u^{\mu}\varepsilon_{\alpha\beta\nu}-2 Y W \epsilon_{\alpha\beta}^{\;\;\;\;\mu}u_{\nu}+2(\dot{W}-H W)\varepsilon^{\mu}_{\;\;\nu[\alpha}u_{\beta]}-2 HW \varepsilon^{\mu}_{\;\;\nu\alpha\beta}
	\end{gather}
	Finally, using the identity 
	\beq
	2\varepsilon_{\mu\nu[\beta}u_{\alpha]}=\varepsilon_{\mu\nu\alpha\beta}+2\varepsilon_{\alpha\beta[\mu}u_{\nu]}
	\eeq
	we obtain the stated result $(\ref{Riemann})$.
\end{proof}
Some comments are now in order.

	\textbf{Comment 1.} From the above expression of the Riemann tensor we see that in a Cosmological $L_{n}$ the latter is completely specified by $6$ time functions for $n=4$ and by $3$ functions for $n\neq 4$. Indeed, this can be easily verified  by constructing all the possible independent invariants with the projection and the $n$-velocity and we trivially find
		\beq
	u^{\alpha}u^{\mu}h^{\nu\beta}R_{\mu\nu\alpha\beta}=-(n-1)\Big[ \dot{X}+H(X+Z+V)+X(Z+V)+(\dot{H}+H^{2})\Big]
	\eeq
	\beq
	u^{\alpha}u^{\nu}h^{\mu\beta}R_{\mu\nu\alpha\beta}=-(n-1)\Big[ \dot{Y}+H(Y+Z+V)-Y(Z+V)-(\dot{H}+H^{2})\Big]
	\eeq
	\beq
	h^{\mu\alpha}h^{\nu\beta}R_{\mu\nu\alpha\beta}=(n-1)(n-2)\Big[ H(X-Y)-X Y-W^{2}+H^{2}\Big]
	\eeq
	\beq
	h^{\mu\beta}h^{\nu\alpha}R_{\mu\nu\alpha\beta}=-(n-1)(n-2)\Big[ H(X-Y)-X Y-W^{2}+H^{2}\Big]
	\eeq
	\beq
	u^{\mu}\varepsilon^{\nu\alpha\beta}R_{\mu\nu\alpha\beta}=2\cdot 3! (H+X)W
	\eeq
	\beq
	u^{\nu}\varepsilon^{\mu\alpha\beta}R_{\mu\nu\alpha\beta}=2 \cdot 3! (Y-H)W
	\eeq
	\beq
	u^{\alpha}\varepsilon^{\mu\nu\beta}R_{\mu\nu\alpha\beta}=(\dot{W}+H W)3!
	\eeq
	\beq
	\varepsilon^{\mu\nu\alpha\beta}R_{\mu\nu\alpha\beta}=2 \cdot 3! \Big[ (\dot{W}+3H W)+W(X-Y) \Big]
	\eeq
	from which we conclude that out of the above $8$ invariant quantities, the third and fourth are the same (with a sign difference) and the last term is a linear combination of the three invariants appearing above it. As a result in a Cosmological $L_{n}$ the Riemann tensor is specified by exactly $6$ functions for $n=4$. This result seems to be in contradiction with \cite{minkevich1998isotropic} where the description of the Riemann tensor in $4$-dimensions was attributed only to $3$ time functions. Probably this contradiction arises because in \cite{minkevich1998isotropic} the pseudo-scalar mode $W$ was constrained to vanish. This is certainly so for $n\neq 4$ but not for $n=4$. For $n\neq 4$ the pseudo-scalar mode disappears and then and only then the Riemann tensor is specified by three functions alone\footnote{Obviously this is so because for $W=0$ the last $4$ terms of the list of invariants vanish identically.}. Come to think about it the appearance of these $6$ degrees of freedom does make sense. The reasoning goes as follows. In $4$ dimensions the Cosmological $L_{n}$ as given by the metric ($\ref{metric}$) and the connection ($\ref{connect}$) is completely specified by $6$ time functions, one coming from the metric (the scale factor) and $5$ from the connection (the distortion variables). These $6$ degrees of freedom are then rearranged in a certain way and produce the  $6$ degrees of freedom that appear in the expression for the Riemann tensor given above.

	\textbf{Comment 2.} The last $3$ terms in (\ref{Riemann}) disappear when $W=0$. That is, these terms arise due to torsion and more precisely due to the totally antisymmetric part of the latter which exists only for $n=4$. Subsequently, for the vanishing torsion subsector, the curvature is specified by three functions only.
	
	Let us also give a list of some essential geometrical objects that are derived from the above expression of the Riemann tensor.
	\begin{itemize}
		\item 	\underline{Ricci Tensor}
		\begin{gather}
		R_{\mu\nu}= \Big[\dot{X}+H\Big( (n-1)X-(n-2)Y+(Z+V)\Big) -(n-2)X Y+X(Z+V)-2 W^{2}+\dot{H}+(n-1)H^{2}\Big] h_{\mu\nu}\nonumber \\
		+\Big[ \dot{Y}+H(Y+Z+V)-Y(Z+V)-(\dot{H}+H^{2})\Big](n-1)u_{\mu}u_{\nu}
		\end{gather}
			\item 	\underline{Homothetic Curvature}
		\begin{gather}
	\hat{R}_{\mu\nu} \equiv 0
		\end{gather}
		\item 	\underline{Co-Ricci}
		\begin{gather}
	\breve{R}_{\mu\nu}=\Big[ \dot{Y}+H \Big(-(n-2)X+(n-1)Y+(Z+V) \Big)+(n-2)X Y -Y (Z+V) + 2 W^{2} -\dot{H}-(n-1)\dot{H}^{2}\Big]h_{\mu\nu} \nonumber \\
	+\Big[ \dot{X}+H (X+Z+V)+X (Z+V)+(\dot{H}+H^{2}) \Big] (n-1)u_{\mu}u_{\nu}
	\end{gather}
		\item \underline{Ricci Scalar}
		\begin{gather}
	R=(n-1)\Big[(\dot{X}-\dot{Y})+(n-1) H (X-Y) +(X+Y)(Z+V)-(n-2) X Y - 2 W^{2}+2 \dot{H}+n H^{2} \Big]
		\end{gather}
		
		\item \underline{Ricci+co-Ricci}	
		\begin{gather}
		(R_{\mu\nu}+	\breve{R}_{\mu\nu})=\Big[ (\dot{X}+\dot{Y})+H (X+Y)+2H (Z+V) +(X-Y)(Z+V) \Big]\Big(h_{\mu\nu}+(n-1)u_{\mu}u_{\nu}\Big)
		\end{gather}
		
	\item \underline{Ricci-co-Ricci}
	
		\begin{gather}
		(R_{\mu\nu}-	\breve{R}_{\mu\nu})=\Big[ (\dot{X}-\dot{Y})+(2n-3) H (X-Y) +(X+Y)(Z+V) -2(n-2) X Y -4 W^{2}+2 \dot{H}+2 (n-1)H^{2}\Big]h_{\mu\nu}\nonumber \\
		-\Big[ (\dot{X}-\dot{Y})+ H (X-Y)+ (X+Y)(Z+V)+2 (\dot{H}+H^{2})\Big](n-1)u_{\mu}u_{\nu}
		\end{gather}
		
		\item \underline{Generalized Einstein Tensor}\footnote{The definition we adopt for the Einstein tensor comes from the contraction of the generalized Riemann double dual. Our definition agrees with the one used in \cite{aoki2019scalar}. We should note however that this generalized version has neither properties of the usual Einstein tensor, that is the generalized Einstein tensor is, in general, neither symmetric nor covariantly conserved.}
		
		\begin{gather}
		G_{\mu\nu}:=\frac{(n-3)}{2}\Big( R_{\mu\nu}-\breve{R}_{\mu\nu}-R g_{\mu\nu}\Big)= \nonumber \\
		=\frac{(n-2)(n-3)}{2}\Big[ -\Big( (\dot{X}-\dot{Y})+(n-2)H(X-Y)-(X+Y)(Z+V)+(n-3)X Y-W^{2}+2 \dot{H}-H^{2} \Big) h_{\mu\nu}\nonumber \\
		+\Big( H(X-Y)-X Y-W^{2}+H^{2}\Big)(n-1)u_{\mu}u_{\nu} \Big]
		\end{gather}

		\item \underline{Right dual}
		\begin{gather}
		R_{\mu\nu\alpha\beta}\varepsilon^{\alpha\beta\kappa\lambda}	=\Big[ \dot{X}+H(X+Z+V)+X(Z+V)+(\dot{H}+H^{2})\Big] 2 u_{\mu}\varepsilon_{\nu}^{\;\;\kappa\lambda}\nonumber \\+\Big[ \dot{Y}+H(Y+Z+V)-Y(Z+V)-(\dot{H}+H^{2})\Big]2 u_{\nu}\varepsilon_{\mu}^{\;\;\kappa\lambda}
		\nonumber \\+\Big[H^{2}+ H(X-Y)-X Y-W^{2}\Big] 2 \Big( \varepsilon_{\mu\nu}^{\;\;\;\;\kappa\lambda}+u_{\nu}\varepsilon_{\mu}^{\;\;\kappa\lambda}-u_{\mu}\varepsilon_{\nu}^{\;\;\kappa\lambda}\Big)+8 XW u_{\mu}\delta_{\nu}^{[\kappa}u^{\lambda]}+8  Y W u_{\nu}\delta_{\mu}^{[\kappa}u^{\lambda]}\nonumber \\
		+4(\dot{W}-H W)h^{\kappa}_{\;\;[\mu}h_{\nu]}^{\;\;\;\lambda}+8  HW \delta^{[\kappa}_{\mu}\delta^{\lambda]}_{\nu}
		\end{gather}
		\item \underline{Symmetric Part of Riemann}
		\begin{gather}
		R_{(\mu\nu)\alpha\beta}=\Big[ (\dot{X}+\dot{Y})+H(X+Y)+2 H(Z+V)+(X-Y)(Z+V)\Big] 2 u_{(\mu}h_{\nu)[\alpha}u_{\beta]}-2 W(X+Y)\varepsilon_{\alpha\beta(\mu}u_{\nu)}
		\end{gather}
		
		\item \underline{Totally Antisymmetric Riemann}\footnote{In the literature this term is sometimes called the Hojman or more frequently the Holst term.}
		\beq
		\varepsilon^{\mu\nu\alpha\beta}R_{\mu\nu\alpha\beta}=12\Big[ (\dot{W}+3 H W)+W ( X-Y) \Big]
		\eeq
			\end{itemize} 
		Of course the above list could go on and on. Here our intention was to present the quantities that play an  actual role in applications. For completeness let us also present the form of the Riemann tensor of the characteristic cases of vanishing non-metricity (Riemann-Cartan Space) and vanishing torsion  (Symmetric Space). For transparency in each case below we express the distortion variables in terms of torsion and non-metric degrees of freedom respectively. We have:
		\subsection{ \underline{Riemann-Cartan Space (vanishing non-metricity)}}
		
		\begin{gather}
		R_{\mu\nu\alpha\beta}	=\Big[ (\dot{H}- 2 \dot{\Phi})+H(H-2 \Phi)\Big]4 u_{[\mu}h_{\nu][\alpha}u_{\beta]}
			+\Big[ (H-2 \Phi)^{2}-P^{2}\Big] 2 h_{\mu[\alpha}h_{\beta]\nu}\nonumber\\
		+\Big[ \dot{P}-P(H-4 \Phi)\Big]2 \varepsilon_{\alpha\beta[\nu}u_{\mu]}-(\dot{P}+HP)\varepsilon_{\mu\nu\alpha\beta} 
		\end{gather}

				\subsection{ \underline{Symmetric Space (vanishing torsion)}}
			\begin{gather}
		R_{\mu\nu\alpha\beta}	=\left[ \Big(\dot{H}- \frac{\dot{A}}{2} \Big)+H\Big(H-\frac{A}{2}\Big)\right]4 u_{[\mu}h_{\nu][\alpha}u_{\beta]}+\left[ \dot{B}+H B +\Big( H-\frac{A}{2}\Big)(A+C) \right]2 u_{(\mu}h_{\nu)[\alpha}u_{\beta]}  \nonumber \\
		+\Big( H-\frac{A}{2} \Big) \left[\frac{B}{2}+\Big( H-\frac{A}{2} \Big) \right] 2 h_{\mu[\alpha}h_{\beta]\nu}
		\end{gather}

	Note: If $P=0$ in the former case and $B=0$, $A=-C$ in the latter case the two resulting Riemann tensors are related through the duality relation $ 2 \Phi \leftrightarrow A/2$. This is the known duality which was reported earlier in the literature (see \cite{iosifidis2019torsion}) and holds true for specific subspaces of torsion/non-metricity.

	\section{Gauss-Bonnet Density}
	
	Having the form of the Riemann tensor, a highly interesting object one can compute is the so-called Gauss-Bonnet term. Of course in metric Theories its form is well-known to be given by
	\beq
		\mathcal{G}=\widetilde{R}_{\mu\nu\alpha\beta}\widetilde{R}^{\alpha\beta\mu\nu}-4\widetilde{R}_{\mu\nu}\widetilde{R}^{\mu\nu}+\widetilde{R}^{2}
	\eeq
	and subsequently the Gauss-Bonnet density $\sqrt{-g}	\mathcal{\hat{G}}$ is known to be a topological invariant (i.e. independent of the metric) in $4$ dimensions and as a total divergence does not contribute to the field equations when included in a Gravity Theory. It is interesting now, to explore what the generalized Gauss-Bonnet term looks like in this non-Riemannian Cosmological background we are considering here, given the form $(\ref{Riemann})$. As a generalized version of the above Riemannian form, we adopt the definition for the generalized Gauss-Bonnet term as given by 
	\beq
	\mathcal{G}:= R_{\mu\nu\alpha\beta}R^{\alpha\beta\mu\nu}-(R_{\mu\nu}-\breve{R}_{\mu\nu})(R^{\nu\mu}-\breve{R}^{\nu\mu})+R^{2} \label{GB}
	\eeq
	Our definition agrees with the one considered in both \cite{babourova1996gauss} and \cite{janssen2019topological}\footnote{Note that in \cite{jimenez2016spacetimes,beltran2017modified} a different definition of the Gauss-Bonnet term was considered. There, a linear combination of quadratic curvature invariants was considered with the demand that in the Riemannian limit this combination will yield the Riemannian Gauss-Bonnet term. } and follows from the contraction of  the Riemann tensor with the Riemann double dual after expanding the generalized Kronecker deltas. Let us also stress out that the inclusion of the aforementioned term in Metric-Affine Gravitational Theories is interesting for various reasons. For instance, this term is projectively invariant\footnote{This aspect is crucial for constructing healthy Theories \cite{aoki2019scalar,BeltranJimenez:2019acz}. } and also this exact linear combination of Riemann squared terms is free of ghosts \cite{jimenez2016spacetimes},  a feature that is not true for general quadratic curvature Theories (see \cite{BeltranJimenez:2019acz}). In addition, this definition is valid for any dimension and $(\ref{GB})$ becomes topological as soon as the connection is reduced to be Levi-Civita.
	
	From the generalized Gauss-Bonnet term we can then define the Gauss-Bonnet density
	\beq
		\mathcal{\hat{G}}:=\sqrt{-g}	\mathcal{G} \label{GBD}
	\eeq
	It is also known that the latter density is topological for Cartan-Weyl spaces \cite{babourova1996gauss} but it fails to be so when the full non-metricity is included \cite{janssen2019topological}. Let us now compute the generalized Gauss-Bonnet term for a Cosmological $L_{n}$. We have the following result.

	\begin{ftheo}
		Consider the generalized Gauss-Bonnet term as given in ($\ref{GB}$). In a Cosmological $L_{n}$ the latter assumes the form 
\begin{gather}
\mathcal{G}=(n-1)(n-2)(n-3)\Big[(n-4)(H^{2}+HX-HY-X Y-W^{2})^{2}\nonumber\\
+ 2 \Big( (\dot{X}-\dot{Y})+ H (X-Y)+ (X+Y)(Z+V)+2 (\dot{H}+H^{2})\Big)(H^{2}+HX-HY-X Y-W^{2})\nonumber \\-4 (\dot{W}+H W)(2HW +WX -WY) \Big] \label{GBCosmo}
\end{gather}
	\end{ftheo}
	\begin{proof}
		Starting with the covariant form ($\ref{Riemann}$) we carefully compute each term appearing in ($\ref{GB}$). We find
			\begin{gather}
		R_{\mu\nu\alpha\beta}R^{\alpha\beta\mu\nu}=(n-1)\Big(  (\dot{X}-\dot{Y})+ H (X-Y)+ (X+Y)(Z+V)+2 (\dot{H}+H^{2}) \Big)^{2}\nonumber \\
		+2(n-1)(n-2)\Big(H^{2}+H(X-Y)-X Y-W^{2}\Big)^{2}\nonumber \\
		-4 ! (\dot{W}W+H W^{2})\Big( 2H+(X-Y)\Big)
		\end{gather}
		\begin{gather}
		(R_{\mu\nu}-	\breve{R}_{\mu\nu})(R^{\nu\mu}-	\breve{R}^{\nu\mu})=(n-1)\Big[(n-1)\Big( (\dot{X}-\dot{Y})+ H (X-Y)+ (X+Y)(Z+V)+2 (\dot{H}+H^{2}) \Big)^{2}\nonumber \\
		+ \Big( (\dot{X}-\dot{Y})+ H (X-Y)+ (X+Y)(Z+V)+2 (\dot{H}+H^{2})+2(n-2)(H^{2}+HX-HY-X Y-W^{2}) \Big)^{2}
		\Big]
		\end{gather}
		\begin{gather}
		R^{2}=(n-1)^{2}\Big( (\dot{X}-\dot{Y})+ H (X-Y)+ (X+Y)(Z+V)+2 (\dot{H}+H^{2})
		\nonumber \\
		+(n-2)(H^{2}+H(X-Y)-X Y - W^{2})\Big)^{2}
		\end{gather}
		Of course the next step is to now substitute all of these back to  equation ($\ref{GB}$). Even though the situation is seemingly complicated it can be greatly simplified by observing that there are certain combinations appearing. To be more specific, defining
			\beq
		\lambda:=	 (\dot{X}-\dot{Y})+ H (X-Y)+ (X+Y)(Z+V)+2 (\dot{H}+H^{2}) \label{l}
		\eeq
		and
		\beq
		\xi:=H^{2}+H(X-Y)-X Y \label{xi}
		\eeq
	the above terms simplify considerably (see Appendix). 	With this observation, and plugging everything into ($\ref{GB}$) after some algebra we arrive at
			\beq
		\mathcal{\hat{G}}=(n-1)(n-2)(n-3)\Big[(n-4)(\xi-W^{2})^{2}+ 2 \lambda(\xi-W^{2})-4 (\dot{W}+H W)(2HW +WX -WY) \Big]
		\eeq
		Note that the $\lambda^{2}$ terms have cancelled out in the final expression. Then
	 by substituting in the latter equation the definitions of $\lambda$ and $\xi$ we arrive at the stated result ($\ref{GBCosmo}$).

	\end{proof}
	 An immediate comment follows from the above result.
	 
\textbf{Comment 3.}
	In a Cosmological $L_{n}$ the generalized Gauss-Bonnet term ($\ref{GB}$) identically vanishes when the number of dimensions is $n=1,2$ or $3$.

Now, one may ask what happens for $n=4$ which is the case of our physical spacetime and compare it with the standard metric case. Of course in the latter case, the Gauss-Bonnet density is a total derivative and as such it does not contribute to the field equations of the Theory when $n=4$. It is interesting to see what happens with the generalized Gauss-Bonnet density here. It is also known that the latter density continues to be a total derivative in Weyl Cartan spaces (see ref. \cite{babourova1996gauss}) but it fails to be so when the full non-metricity is included \cite{babourova1997pontryagin} (see also \cite{janssen2019topological}). It is then natural to ask, are there any extended geometries more general then Cartan-Weyl, for which the Gauss-Bonnet density ($\ref{GBD}$) continues to be a total derivative. As we show below the answer to this question is affirmative, for the Cosmological setting we are considering here. We have the following result.
	
		\begin{ftheo}
	For the restricted $B=0$ subsector of a Cosmological $L_{n}$ the generalized Gauss-Bonnet density is a total derivative in $4$ dimensions. In particular, for $B=0$  and $n=4$ it holds that
		\begin{gather}
\mathcal{\hat{G}}=\sqrt{-g}\mathcal{G}= \frac{d}{dt}\Big[8 a^{3}x\Big(x^{2}-3 W^{2}\Big) \Big] 
		\end{gather}
		where 
		\beq
		x=H-Y
		\eeq
	\end{ftheo}
	\begin{proof}
		We start by a making a trivial rearrangement in the expressions we have for $\lambda$ and $\xi$ which with the help of ($\ref{XY}$) recast to
		\beq
		\lambda=2(\dot{H}-\dot{Y})+2 H(H-Y)+\frac{1}{2}(\dot{B}+HB)+\frac{B}{2}(Z+V)
		\eeq
		\beq
		\xi=(H-Y)^{2}+\frac{B}{2}(H-Y)
		\eeq
		Defining 
		\beq
		x:=H-Y
		\eeq
		 and setting $B=0$ they read
		\beq
		\lambda=2(\dot{x}+H x)\;\;, \;\;\; \xi=x^{2}
		\eeq
		Then, if we take these expressions and substitute them back in ($\ref{GBCosmo}$) for $n=4$, after some algebra we recognise that these terms form the derivative of a product and finally we get (see Appendix)
		\begin{gather}
		\mathcal{\hat{G}}=\sqrt{-g}\mathcal{G}= \frac{d}{dt}\Big[8 a^{3}x\Big(x^{2}-3 W^{2}\Big) \Big] 
		\end{gather}
		as stated.
	\end{proof}
	\textbf{Comment 4.} Let us note that the geometry with $B=0$ we considered above is a generalization of the Weyl-Cartan geometry where the non-metricity apart from the usual Weyl part has another degree of freedom\footnote{This can be easily seen from ($\ref{Qcosm}$). Recall that for Weyl non-metricity one has $B=0$ and $A=-C$. Here we have only $B=0$ without the latter constraint (i.e no relation between $A$ and $C$), which represents a generalization of the Weyl spacetime.}. Therefore we conclude that the generalized Gauss-Bonnet density is a total derivative term for this extended Cartan-Weyl geometry in a Cosmological setting.
	
	We shall now prove that for $B\neq 0$ the density  ($\ref{GBCosmo}$) is not a total derivative in $4$ dimensions. Leaving  $B \neq 0$ and carrying out a similar analysis as above, after spotting   the various appearances of total derivatives  we finally find (see Appendix for details)
	\beq
	\mathcal{\hat{G}}=\frac{d}{dt}\Big[ a^{3}\Big(8x(x^{2}-3 W^{2})+6B(x^{2}-W^{2})\Big) \Big] +\frac{a^{3}x B}{4}(\dot{B}+HB)+\frac{a^{3}}{2}B(Z+V)\Big( x^{2}+\frac{B}{2}x- W^{2}\Big) \label{GBCo2}
	\eeq
	From which we see that the first term  of the latter expression, is obviously a total derivative derivative but it is impossible to express the last two combinations as total differentials. We see therefore that the presence of $B$ has spoiled the total derivative nature of the generalized Gauss-Bonnet density. It is also interesting to note that for higher dimensions ($n>4$) the Gauss-Bonnet density cannot (in general) be written as a total differential even in the restricted $B=0$ geometry because of the very presence of the term\footnote{Note that for $n\neq 4$ the $W$ term disappears as we have already discussed in the introduction.} $(n-4)\xi^{2}$. Collecting all the aforementioned results we have the following corollary.
	\begin{corollary}
		In a Cosmological $L_{n}$ the generalized Gauss-Bonnet density ($\ref{GBD}$)
		\begin{itemize}
			\item Identically vanishes for $n=1,2$ and $3$, where $n$ denotes the number of dimensions.
			\item For $n=4$ it is not in a general a total derivative. For the reduced space $B=0$, which serves as a certain generalization of Weyl-Cartan space, the latter density becomes indeed a total differential.
			\item For $n>4$ the Gauss-Bonnet density is not, in general, a total derivative term even for the restricted case $B=0$. 
		\end{itemize}
		\end{corollary}
	 
	 \section{On Applications: Gauss-Bonnet Gravity in Metric-Affine Formalism}
	 As an application of our derived results,	let us start consider the usual Einstein-Hilbert action and add to it the above defined Gauss-Bonnet term. As we have already proved since our affine connection is a priori  generic, the Gauss-Bonnet term will not be a total derivative even in $4$ dimensions. It is interesting then, to study the Cosmology of this Theory. So our starting point will be the action
	 	\beq
	 	S[g,\Gamma]=\frac{1}{2 \kappa}\int d^{n}x\sqrt{-g} \Big( R+\lambda \hat{G} \Big) 
	 	\eeq
	 	where $\lambda$ is a parameter. We leave the spacetime dimension $n$ unspecified at this point but we are going to specialize to $n=4$ shortly. Note that the Gauss-Bonnet term is projectively invariant in any dimension and is also conformally invariant only in $4$ dimensions. So in $4$ dimensions the Gauss-Bonnet term is both projectively and conformally invariant. Let us now find the field equations of the Theory. Variation with respect to the metric gives
	 	\beq
	 	R_{(\mu\nu)}-\frac{R}{2}g_{\mu\nu}+\lambda \Big( 4 G^{\alpha\beta}_{\;\;\;\;(\mu|\lambda}R_{\alpha\beta|\nu)}^{\;\;\;\;\;\;\;\;\lambda} +2 G^{\lambda}_{\;\;(\nu|\alpha\beta|}R_{\mu)\lambda}^{\;\;\;\;\;\alpha\beta}-\frac{1}{2}g_{\mu\nu}\hat{G} \Big)=0
	 	\eeq
	 	Contracting the latter with the metric, we obtain
	 	\beq
	 	(2-n)R+\lambda(4-n)\hat{G}=0 \label{RG}
	 	\eeq
	 	from which we conclude that in $4-dim$ the contribution from the Gauss-Bonnet term drops out. Of course this is to be expected, since as we pointed out, in this dimensionality the Gauss-Bonnet term is conformally invariant, and therefore its variation with respect to the metric produces a traceless tensor. To close the system of field equations we also vary with respect to the connection, to get
	 	\beq
	 	P_{\lambda}^{\;\;\;\mu\nu}-\lambda \left( 2\frac{\nabla_{\alpha}(\sqrt{-g}G_{\lambda}^{\;\;\mu\nu\alpha})}{\sqrt{-g}}+4 G_{\lambda}^{\;\;\mu\nu\alpha}S_{\alpha}+ 2 G_{\lambda}^{\;\;\mu\alpha\beta}S_{\alpha\beta}^{\;\;\;\;\nu} \right)=0 \label{P}
	 	\eeq
	 	where $P_{\lambda}^{\;\;\;\mu\nu}$ is the so-called Palatini tensor and is explicitly given by
	 	\beq
	 	P_{\lambda}^{\;\;\;\mu\nu}=-\frac{\nabla_{\lambda}(\sqrt{-g}g^{\mu\nu})}{\sqrt{-g}}+\frac{\nabla_{\sigma}(\sqrt{-g}g^{\mu\sigma})\delta^{\nu}_{\lambda}}{\sqrt{-g}} \\
	 	+2(S_{\lambda}g^{\mu\nu}-S^{\mu}\delta_{\lambda}^{\nu}+g^{\mu\sigma}S_{\sigma\lambda}^{\;\;\;\;\nu})
	 	\eeq
	 	Now, due to the fact that both the Ricci scalar and the Gauss-Bonnet term are projectively invariant, equation $(\ref{P})$ is identically traceless when contracted in $\mu=\lambda$. Setting now $n=4$ in $(\ref{RG})$ we conclude that $R=0$. Expanding the latter equation and considering a homogeneous FLRW background we arrive at
	 	\begin{gather}
	 \frac{\ddot{a}}{a}+\left(\frac{\dot{a}}{a}\right)^{2}+ 4 \Phi^{2}-P^{2}+\frac{1}{8}\Big( 2 A^{2}+B(C-A) \Big) +\Phi(2 A-B)+\dot{f}+3 H f=0 \;\;, \;\; 	 f:=\frac{1}{2}\left[  \frac{B}{2}-A 	-4 \Phi \right] 
	 \end{gather}
	 which is a variant of the modified Friedmann equation. As for the acceleration equation, the form of the latter for non-Riemannian Universes has been derived in \cite{iosifidis2020cosmic} and reads
	 	\begin{gather}
	 \frac{\ddot{a}}{a}=-\frac{1}{3}R_{\mu\nu}u^{\mu}u^{\nu}+2\left( \frac{\dot{a}}{a} \right)\Phi +2\dot{\Phi} 
	 +\left( \frac{\dot{a}}{a} \right)\left(A+\frac{C}{2}\right)+\frac{\dot{A}}{2}-\frac{A^{2}}{2}-\frac{1}{2}AC  
	 -2A\Phi-2C \Phi\label{accel}
	 \end{gather}
	 One could then proceed by contracting  the metric field equations with $u^{\mu}u^{\nu}$ in order to eliminate the first term ($R_{\mu\nu}u^{\mu}u^{\nu}$) and express everything in terms of the scale factor and the torsion and non-metricity variables. This results in a fairly complicated expression which we refrain from presenting  here since it goes beyond the scope of the present treatment. From the above equations we see that both torsion and non-metricity have interesting and highly non-trivial effects on the behaviour of the expansion.
	 
	\section{Conclusions}
	
	We have derived the covariant form of the generalized Riemann tensor (i.e. including also torsion and non-metricity) for non-Riemannian Cosmological spacetimes. This is the most general covariant expression for the Riemann tensor that is compatible with the Cosmological Principle. In deriving this expression we used the earlier reported covariant forms of torsion and non-metricity for homogeneous Cosmological settings. Our result shows that in a Friedman-like Cosmological setting, the curvature is completely specified by $6$ time functions when the spacetime dimension is equal to $4$ while for $n \neq 4$  the latter possesses only $3$ degrees of freedom. It goes without saying, that the aforementioned covariant form is an indispensable ingredient for the study of Cosmological aspects of Metric-Affine Gravity. Therefore, our result is directly applicable and serves as a necessary ingredient in exploring the role of non-Riemannian degrees of freedom  in Cosmology.
	
	Having obtained the covariant form of the Riemann tensor in a Cosmological $L_{n}$ we then went on and computed some  of the basic objects that appear in applications, such as the Ricci tensor, Co-Ricci tensor, Ricci scalar etc. With the help of these relations we then proceeded in computing the generalized Gauss-Bonnet  density. As expected  our results show that the presence of the full non-metricity makes it impossible to express the latter density as a total differential in general. However, we proved that by constraining slightly the geometry by setting one of the non-metricity degrees of freedom to zero, the generalized version of the Gauss-Bonnet density becomes indeed a total derivative. Note that it was known in the literature that for Cartan-Weyl geometries (i.e. arbitrary torsion but Weyl non-metricity) the latter is indeed always a total differential \cite{babourova1996gauss}. Here we have shown that in a Cosmological setup, the Gauss-Bonnet density is a total differential for geometries for which the non-metricity is more general from Weyl. Of course our result holds true for a Cosmological setup but this should not lead us to the conclusion that this would be so for any background. Here we proved that this is so only for highly symmetric FLRW backgrounds.	
	It would be interesting, however, to see under what circumstances the Gauss-Bonnet term we presented here is a total derivative for the more general family of spherically symmetric Metric-Affine geometries as derived in \cite{hohmann2020metric}\footnote{Possibly, it would also be interesting to find the Gauss-Bonnet density analogues for the symmetric teleparallel \cite{nester1998symmetric,jimenez2018teleparallel} and general teleparallel \cite{jimenez2019general} Theories as was found for metric teleparallelism in \cite{kofinas2014teleparallel}. }. Finally, it would be interesting to find the most general structure of the non-Riemannian geometries for which (\ref{GB}) is a total derivative irrespective of the background.

	\section{Acknowledgments}
I would like to thank Christos Charmousis, Alejandro Jimenez Cano and Konstantinos Palikaris for useful discussions.
	This research is co-financed by Greece and the European Union (European Social Fund- ESF) through the
	Operational Programme "Human Resources Development, Education and Lifelong Learning" in the context
	of the project "Reinforcement of Postdoctoral Researchers - 2
	nd Cycle" (MIS-5033021), implemented by the
	State Scholarships Foundation (IKY).

	\appendix

	\section{Supplementary material for the proofs}
\subsection{Gauss-Bonnet density for $L_{n}$ Cosmology}	Below we give some more details regarding the derivation of the Gauss-Bonnet term $(\ref{GB})$. In computing the invariants entering the Gauss-Bonnet term the following relations were quite helpful
\beq
u^{\mu}R_{\mu\nu\alpha\beta}=-\Big[ \dot{X}+H(X+Z+V)+X(Z+V)+(\dot{H}+H^{2})\Big] 2 h_{\nu[\alpha}u_{\beta]}+2 W(X+H)\varepsilon_{\alpha\beta\nu}
\eeq
	\beq
	u^{\nu}R_{\mu\nu\alpha\beta}=-\Big[ \dot{Y}+H(Y+Z+V)-Y(Z+V)-(\dot{H}+H^{2})\Big]2 h_{\mu[\alpha}u_{\beta]}+2 W(Y-H)\varepsilon_{\mu\alpha\beta}
	\eeq
	\begin{gather}
		u^{\alpha}R_{\mu\nu\alpha\beta}=\Big[ \dot{X}+H(X+Z+V)+X(Z+V)+(\dot{H}+H^{2})\Big]u_{\mu}h_{\nu\beta}+(\dot{W}+H W)\varepsilon_{\mu\nu\beta}
		\nonumber \\+\Big[ \dot{Y}+H(Y+Z+V)-Y(Z+V)-(\dot{H}+H^{2})\Big]u_{\nu}h_{\mu\beta}
			\end{gather}
	\beq
	h^{\mu\alpha}R_{\mu\nu\alpha\beta}=(n-1)\Big[ \dot{Y}+H(Y+Z+V)-Y(Z+V)-(\dot{H}+H^{2})\Big] u_{\nu}u_{\beta}+(n-2)\Big[ H(X-Y)-X Y-W^{2}+H^{2}\Big]h_{\nu\beta}
	\eeq
		\beq
	h^{\nu\alpha}R_{\mu\nu\alpha\beta}=(n-1)\Big[ \dot{X}+H(X+Z+V)+X(Z+V)+(\dot{H}+H^{2})\Big] u_{\mu}u_{\beta}-(n-2)\Big[ H(X-Y)-X Y-W^{2}+H^{2}\Big]h_{\mu\beta}
	\eeq
	\beq
	u^{\mu}u^{\alpha}R_{\mu\nu\alpha\beta}=-\Big[ \dot{X}+H(X+Z+V)+X(Z+V)+(\dot{H}+H^{2})\Big]h_{\nu\beta}
	\eeq
		\beq
	u^{\nu}u^{\alpha}R_{\mu\nu\alpha\beta}=-\Big[ \dot{Y}+H(Y+Z+V)-Y(Z+V)-(\dot{H}+H^{2})\Big] h_{\mu\beta}
	\eeq
	\beq
	h^{\mu\nu}	R_{\mu\nu\alpha\beta}=0=u^{\mu}u^{\nu}	R_{\mu\nu\alpha\beta}
	\eeq
	Then with the abbreviations ($\ref{l}$) and $(\ref{xi})$ we find for each piece
		\beq
	R_{\mu\nu\alpha\beta}R^{\alpha\beta\mu\nu}=(n-1)\lambda^{2}+2(n-1)(n-2)\Big(\xi-W^{2}\Big)^{2}-4 ! (\dot{W}W+H W^{2})\Big( 2H+(X-Y)\Big)
	\eeq
	\beq
	(R_{\mu\nu}-	\breve{R}_{\mu\nu})(R^{\nu\mu}-	\breve{R}^{\nu\mu})=(n-1)\Big[ \Big(\lambda+2(n-2)(\xi-W^{2}) \Big)^{2}+(n-1)\lambda^{2}\Big]
	\eeq
	\beq
	R^{2}=(n-1)^{2}\Big( \lambda+(n-2)(\xi - W^{2})\Big)^{2}
	\eeq
	and by substituting the latter back in ($\ref{GB}$) we see that the $\lambda^{2}$ cancel out and we straightforwardly arrive at ($\ref{GBCosmo}$).

	\subsection{Case $n=4$, $B\neq 0$}
Let us now derive the stated form ($\ref{GBCo2}$) for $n=4$ and  the general case $B\neq 0$. For $n=4$ the Gauss-Bonnet density becomes
\beq
\mathcal{\hat{G}}=12\Big[ a^{3}\lambda (\xi-W^{2})-2(\dot{W}W+HW^{2})\Big(2 x +\frac{B}{2}\Big) \Big]
\eeq
Now it is convenient to set
\beq
y=2x +\frac{B}{2}
\eeq
and thus definitions (\ref{l}) and (\ref{xi}) become respectively
\beq
\lambda= (\dot{y}+H y)+\frac{1}{2}(\dot{B}+HB)+\frac{B}{2}(Z+V)
\eeq
\beq
\xi=\frac{1}{4}y^{2}-\frac{1}{16}B^{2}
\eeq
With these the above expression recasts to
\begin{gather}
\frac{1}{12}\mathcal{\hat{G}}=a^{3}(\dot{y}+H y)\xi -a^{3}(\dot{y}+H y)W^{2}+a^{3}\frac{B}{2}(Z+V)(\xi- W^{2})-2y a^{3}(W\dot{W}+H W^{2})
\end{gather}
We observe now that the second and the last term of the above can be added to give the total derivative of a product yielding
\beq
-a^{3}(\dot{y}+H y)W^{2}-2y a^{3}(W\dot{W}+H W^{2})=-a^{2}W^{2}\frac{d}{dt}(a y)-a y \frac{d}{dt}(W^{2}a^{2} )=-\frac{d}{dt}\Big( a^{3}y W^{2}\Big)
\eeq
Substituting also $\xi$ and performing the same technique, after pairing all the terms that form derivatives of products we find
	\beq
\mathcal{\hat{G}}=\frac{d}{dt}\Big[ a^{3}\Big(8x(x^{2}-3 W^{2})+6B(x^{2}-W^{2})\Big) \Big] +\frac{a^{3}x B}{4}(\dot{B}+HB)+\frac{a^{3}}{2}B(Z+V)\Big( x^{2}+\frac{B}{2}x- W^{2}\Big)
\eeq
where $x=H-Y$. Note that in the Riemannian limit ($i.e.$ $B=0=Z=V=W=Y$) the latter reduces to the metric Gauss-Bonnet density
\beq
\mathcal{\hat{G}}=\frac{d}{dt}( 8
\dot{a}^{3} )
\eeq
as expected.
	
	\bibliographystyle{unsrt}
	\bibliography{ref}
	
		\end{document}